\documentclass[12pt]{article}

\usepackage{bm}
\usepackage{amstext}
\usepackage{graphicx}
\usepackage[normalem]{ulem}

\begin{document}

\title{Magnetic properties and concurrence for fluid $^3He$ on kagome lattice   }
\author{ N.S. Ananikian$^1$,  L.N. Ananikyan$^1$   and H.A.
Lazaryan$^2$\\
\begin{footnotesize}\textit{$^1$
A.I. Alikhanyan National Science Laboratory, Alikhanian Br. 2, 0036 Yerevan, Armenia,}\end{footnotesize}\\
\begin{footnotesize}\textit{$^2$
Department of Theoretical Physics, Yerevan State University,
}\end{footnotesize}\\
\begin{footnotesize}\textit{
A. Manoogian 1, 0025 Yerevan, Armenia.}\end{footnotesize}}
\date{\today}
\maketitle

\begin{abstract}
We present the results of magnetic properties and entanglement for
kagome lattice using Heisenberg model with two-, and three-site
exchange interactions in strong magnetic field. Kagome lattice
correspond to the third layer of fluid  $^3He$ absorbed on the
surface of graphite. The magnetic properties and concurrence as a
measure of pairwise thermal entanglement are studied by means of
variational mean-field like treatment based on Gibbs-Bogoliubov
inequality. The system exhibits different magnetic behaviors,
depending on the values of the exchange parameters ($J_2, J_3$). We
have obtained  the magnetization plateaus at low temperatures. The
central theme of the paper is the comparing the entanglement and
magnetic behavior for kagome lattice. We have found that in the
antiferromagnetic region behaviour of the concurrence  coincides
with the magnetization one. \end{abstract}

\section{Introduction}
One can model solid  and fluid $^3He$ films as the systems of almost
localized identical fermions. Since the light mass spin-1/2  $^3He$
atoms are subject to a weak attractive potential, the theoretical
explanation of magnetism is based on the multiple-spin exchange
mechanism \cite{Roger}. An important case is represented by solid
and fluid $^3He$ films absorbed on the surface of graphite
\cite{graffite,graffite2,graffite3} since it is a typical example of
a two-dimensional frustrated quantum-spin system \cite{frustrated}.
The first and second layers of the system form a triangular lattice
\cite{kagome22}, while the third one forms a system of quantum 1/2
spins on a kagome lattice \cite{kagome2}.

The phenomenon of magnetization plateau has been studied during the
past decade both experimentally and theoretically. The plateaus may
be exhibited in the magnetization curves of quantum spin systems at
very low temperatures in case of strong external field.
Magnetization plateaus appear in a wide range of  models on chains,
ladders, hierarchical lattices, theoretically analysed by dynamical,
transfer matrix approaches and  exact diagonalization in clusters
(see Ref.~\cite{platoess}-\cite{platoes13}). In \cite{platoekagome}
dynamical system theory  has been used to study magnetization
plateaus on the kagome chain with two-, three- and six-site exchange
interactions.

Recent years much effort has been put into studying the entanglement
of multipartite systems both qualitatively and quantitatively
\cite{entanglement2,wang}. Entanglement has gained renewed interest
with the development of quantum information science.  Entangled
states constitute a valuable resource in quantum information
processing \cite{entanglement1}. Numerous different methods of
entanglement measuring have been proposed for its quantification
\cite{entanglement0}. In this paper we use concurrence
\cite{wooters} as entanglement measure of the spin-1/2  system.

In the present paper mean-field like approach, based on the Gibbs-Bogoliubov inequality, was used to study entanglement  and  magnetic properties of a kagome lattice \cite{bogolubov}. This
method can also be applied to study thermal entanglement
in many-body systems \cite{bogolubov1,levon}.

The key result of the paper is concentrated on the comparison of
specific (peaks and plateaus) features in magnetization and thermal
entanglement properties in the above mentioned model using
variational mean-field like Gibbs-Bogoliubov inequality.

This paper is organized as follows: in Section 2 we introduce the
Heisenberg model on the kagom´e lattice with two- and three-site
exchange interactions. In Section 3 mean-field like approximation,
based on Gibbs-Bogoliubov inequality, has applied on  kagome
lattice. The magnetic properties of the model are investigated in
Section 4. In Section 5  concurrence  as a measure of entanglement
is studied and compared  with magnetic properties of kagome lattice.
The conclusive remarks are given in Sec. 6.

\section{Heisenberg model Hamiltonian on kagome lattice with two-, and three-site exchanges interactions}
The  Hamiltonian of the Heisenberg model  is \begin{equation}
H = H_{ex}  + H_Z,
\end{equation}
where $H_{ex}$ represent spin  exchanges  and $H_Z$ is responsible for magnetism. The expression for $H_Z$
can be written as\begin{equation}\label{Hz}
H_Z  =  - \sum\limits_i {\frac{\gamma }{2}\hbar \bm{B}\bm{\sigma} _i}
\equiv-\mathrm{h}\ \sum\limits_i {\sigma^z _i
},\end{equation}
where $\gamma$ ,\textbf{B}-- are gyromagnetic ratio and magnetic field.
According to \cite{Roger} the multiple spin exchanges Hamiltonian  can be
written as
\begin{equation}
\label{Hamiltonian23}
H_{ex}  = \mathrm{J}_2 \sum\limits_{\left\langle i,j \right\rangle } {P_{i,j} }  - \mathrm{J}_3 \sum\limits_{\left\langle i,j,k \right\rangle}
{\left( {P_{i,j,k}  + P_{i,j,k}^{ - 1} } \right)+}\ldots,
\end{equation}
\begin{figure}[t]\begin{center}
\includegraphics[width=250pt]{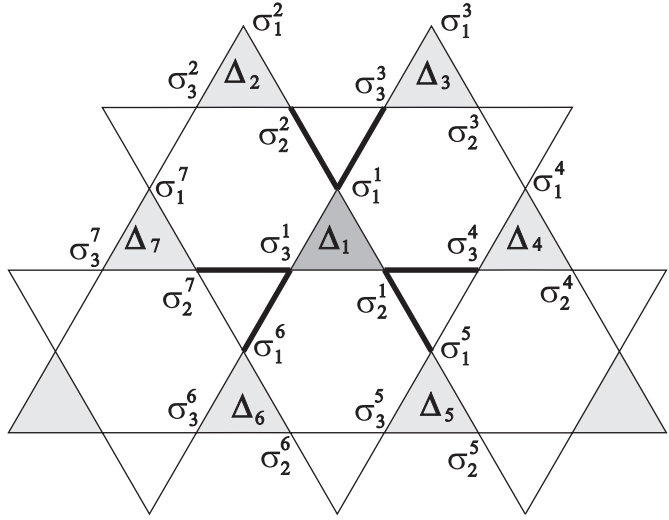}\end{center} \caption{
Kagome lattice. }
\end{figure}where $P_{i,j},P_{i,j,k}$ represent the two-, and three-spin cyclic permutation  operators. The sums are taken over all distinct
 two- and three-cycles. The expression
of pair transposition operator $P_{ij}$ has been given by Dirac\begin{equation}\label{p2}
 P_{i,j}= \frac{1}{2}\left( {1 +  \bm{\sigma} _i
\bm{\sigma} _j } \right)  ,\end{equation} where $\bm{\sigma} _i$ are
the Pauli matrixes, acting on the spin at the $i$-th site. Using this expression
for  $P_{i,k}$  one can find the expressions for the other operators of cyclic
rearrangement, (see \cite{Roger,platoes11}) Here is the expression
for three  spin exchange operator:\begin{equation}
P_{i,j,k}  + P_{i,j,k}^{ - 1} =\frac{1}{2}\left(1+ {\bm{\sigma} _i \bm{\sigma} _j  + \bm{\sigma} _j \bm{\sigma} _k +
\bm{\sigma} _k \bm{\sigma} _i } \right).
\end{equation}

As mentioned above the third layer of $^3He$ system is kagome lattice
(see Fig 1).\ \ In kagome lattice each edge belongs to only one triangle and each site belongs to  two triangles, therefore one can combine first two summations in  (\ref{Hamiltonian23}) and Hamiltonian for kagome lattice can be written in the following form:
\begin{equation}\label{Hkagome}
H =\sum\limits_{Triangles} {\left[\frac{{\mathrm{J}_2-\mathrm{J}_3
}}{2}\left( {\bm{\sigma} _i \bm{\sigma} _j  + \bm{\sigma} _j
\bm{\sigma} _k + \bm{\sigma} _k \bm{\sigma} _i }
\right)-\frac{\mathrm{h}}{2}(\sigma^z _i+\sigma^z _j+\sigma^z
_k)\right]}.
\end{equation}
 According to \cite{kagome22} the effective value of the exchange
parameters $J=J_{2}-2J_{3}$ on triangular lattice, which has been estimated experimentally from susceptibility
and specific-heat data  for solid $^3He$  is
$J=-3.07 mK$. Therefore, the three-site exchange ($J_3$) is dominant on the triangular lattice. This is a consequence of the fact that on triangular lattice, for high densities
 the probability of a triple permutations of  $^3He$
atoms is dominant than a pair one. In the case of fluid  $^3He$, which is
described by kagome lattice, pair exchanges become more probable, since every edge belongs to  one triangle and  one hexagon, in contrast to the triangular lattice for which every edge belongs to two triangles.
Moreover, we did not take into  account the six-site exchange interaction
which is the antiferromagnetic one. Taking into account  above mentioned facts we can consider effective pair exchange permutations ($J_2$) more dominant than  three-site
one ($J_3$).

\section{Basic Gibbs-Bogoliubov mean-field formalism}

Here we apply the variational mean-fi¯eld like treatment based on Gibbs-Bogoliubov
inequality \cite{bogolubov} to solve the Hamiltonian  (\ref{Hkagome}). This implies that the free energy (Helmholtz
potential) of system is
\begin{equation}\label{bogolubovinequality}
F \le F_0  + \left\langle {H - H_0 } \right\rangle _0,
\end{equation}
where $H$ is the real Hamiltonian which describes the system and $H_0$ is the trial one. $F$
and $F_0$ are free energies corresponding to $H$ and $H_0$ respectively and $ \left\langle \dots  \right\rangle _0$ denotes the
thermal average over the ensemble defined by $H_0$. By introducing trial
Hamiltonian for our model (Eq. (\ref{Hkagome}) kagome lattice) containing unknown  variational parameters one can  minimize right hand side of Bogoliubov inequality (\ref{bogolubovinequality}) and get the values of those parameters,

 For antiferromagnetic interactions,  the trial Hamiltonian will consist of two parts describing the two sublattices.
We introduce a trial Hamiltonian $ H_0 $ as a set of noninteracting
clusters (triangles)
  on two sublattices in different external self-consistent  fields:

\begin{equation}\label{trialsum}
H_0  = \sum\limits_{\Delta_i} {H_0^{(i)}  },
\end{equation}
where
\begin{equation}\label{Htrial}
H_0^{(i)} =\lambda \times {\left( {\bm{\sigma}^i_1 \bm{\sigma}^i_2  + \bm{\sigma} ^i_2 \bm{\sigma}^i_3 +
\bm{\sigma}^i_3\bm{\sigma}^i_1 } \right)- \gamma_\upsilon \times\left[ (\sigma^{i}_1)^z+ (\sigma^{i}_2)^z+ (\sigma^{i}_3)^z\right],\nonumber
}
\end{equation}
where $\lambda$ and $\gamma_\upsilon$ variational parameters, and
  $\Delta_i$ labels different noninteracting rectangles (see Fig. 1, grey triangles) and
\begin{equation}\nonumber
\begin{array}{cc}
  \gamma_\upsilon= \gamma_a& \hbox{for sublattice \textbf{(a)}},\\
  \gamma_\upsilon= \gamma_b&\hbox{for sublattice \textbf{(b)}.}\\
\end{array}
\end{equation}
 It should be emphasized
that in trial Hamiltonian spins $\bm{\sigma}^i_k$ of the $\Delta_i$-th triangle do
not interact with the
spins $\bm{\sigma}^{j}_k$ of the $\Delta_{j}$  triangle if $i\neq j$, therefore
these spins are statistically independent. Suppose the real Hamiltonian $H$ (\ref{Hkagome})
can be represented
in the following form
\begin{equation}\label{sum}
H  = \sum\limits_{\Delta_i} {H^{(i)}  },
\end{equation}
where  $H^{(i)}$  is the contribution of
 spins on the single  triangle in real Hamiltonian $H$ and index of summation $\Delta_i$  runs over the different triangles (see Fig. 1, grey triangles). Terms of real Hamiltonian $\bm{\sigma}^i_1 \bm{\sigma}^i_2  + \bm{\sigma} ^i_2 \bm{\sigma}^i_3 +
\bm{\sigma}^i_3\bm{\sigma}^i_1$ must be included in  $H^{(i)}$, but terms
like    $\bm\sigma^{i}_\alpha\bm\sigma^{j}_{\beta}$ (see Fig.~1 solid lines) should be included both in $H^{(i)}$ and $H^{(j)}$. Consequently,  $H^{(i)}$ has
the following form:

\begin{equation}
H^{(i)}=\frac{{\mathrm{J}_2-\mathrm{J}_3
}}{2}\left(\bm\alpha^{(i)}+\sum\limits_{\tau=2,3}^{}\frac{\bm{\sigma}^i_1 \bm{\sigma}^j_\tau}{2}+\sum\limits_{k=1,3}^{}\frac{\bm{\sigma}^i_2 \bm{\sigma}^k_\tau }{2}+\sum\limits_{a=1,2}^{}\frac{\bm{\sigma}^i_3 \bm{\sigma}^l_\tau}{2} \right)-\mathrm{h}\sum\limits_{\tau=1}^{3}(\sigma^{i}_\tau)^z,
\end{equation}
where
  \begin{equation}\label{alphabeta}
  \bm\alpha^{(i)}=\bm{\sigma}^i_1 \bm{\sigma}^i_2  + \bm{\sigma} ^i_2 \bm{\sigma}^i_3 +
\bm{\sigma}^i_3\bm{\sigma}^i_1.
\end{equation}
In the expressions for $H^{(i)}$ we take half of each term $\bm{\sigma}^i_a \bm{\sigma}^j_b$  because this term should be included
in two different triangles.

Inequality (\ref{bogolubovinequality}) can be rewritten now for the
single triangle on each sublattice $(\upsilon)$:
\begin{equation}\label{bogolubovinequality1}
f_\upsilon\  \le (f_0)_\upsilon+
 \left\langle {H^{(i)} - H^{(i)}_0 } \right\rangle _0,
\end{equation}
where $H^{(i)}$ is the real and $H^{(i)}_0$ the trial Hamiltonians
of the triangle, $f_\upsilon$ and $(f_0)_\upsilon$ free energies of the one triangle
on sublattice $(\upsilon)$ defined by $H^{(i)}$ and $H^{(i)}_0$ respectively. By denoting  magnetizations of the sublattices $(a)$ and $(b)$ respectively
$m_a$ and $m_b$ and taking into
account that spins $\sigma^{i}_\tau$ belong to  sublattice $(a)$ and spins
$\sigma^{j,k}_\tau$ belong to  sublattice $(b)$ and fact that
spins $\bm\sigma^{i}_\tau$ and $\bm\sigma^{j,k}_\tau$  \ $(i\neq j,k)$ are
statistically independent we get:  $\left\langle(\sigma^{i}_\tau)^{x,y}\right\rangle=0$,
$m_a\equiv\left\langle(\sigma^{i}_\tau)^{z}\right\rangle/2$, $m_b\equiv\left\langle(\sigma^{j,k}_\tau)^{z}\right\rangle/2$
and $\left\langle\bm\sigma^{i}_\tau\bm
\sigma^{j}_\tau\right\rangle
=\left\langle(\sigma^{i}_a)^{z}\right\rangle\times\left\langle(\sigma^{j}_b)^{z}\right\rangle=2m_a2m_b$. One can rewrite inequality (\ref{bogolubovinequality1})
as follows: \begin{eqnarray}\label{inequality} f_\upsilon\  &\le& (f_0)_\upsilon
+\left(\frac{{\mathrm{J}_2-\mathrm{J}_3 }}{2}-\lambda\right)
\left\langle\bm\alpha\right\rangle_0
+\frac{{\mathrm{J}_2-\mathrm{J}_3
}}{4}6(2m_a 2m_b)-\left(\mathrm{h}-\gamma_\upsilon\right)6m_\upsilon. \nonumber\\
&&\gamma_a=\mathrm{h}-(\mathrm{J}_2-\mathrm{J}_3)
m_b,\quad \gamma_b=\mathrm{h}-(\mathrm{J}_2-\mathrm{J}_3)
m_a.\nonumber
\end{eqnarray}
Minimizing the right hand side  of (\ref{inequality}) in order to $\gamma_a, \gamma_b$ and $\lambda$
and using the fact, that $\displaystyle\frac{{\partial f_0}}{{\partial \lambda}}=\left\langle\bm\alpha\right\rangle_0$
and $\displaystyle\frac{{\partial f_0}}{{\partial \gamma_\upsilon}}=-6m_\upsilon$ we obtain the following values for
the variational parameters:
\begin{eqnarray}\label{gamma2}
 \lambda&=&\frac{{\mathrm{J}_2-\mathrm{J}_3
}}{2},  \nonumber\\
\gamma_a&=&\mathrm{h}-(\mathrm{J}_2-\mathrm{J}_3)
m_b.\nonumber\\
\gamma_b&=&\mathrm{h}-(\mathrm{J}_2-\mathrm{J}_3)
m_a.
\end{eqnarray}

The Hamiltonian $H^{(i)}_0$ was chosen to be
exactly solved.
By diagonalization Hamiltonian  $H^{(i)}_0$ one can find eigenvectors end
eigenvalues of the trial Hamiltonian \cite{wang,levon}.

\section{Magnetic properties}

Here and further  exchange parameters $(J_2,J_3)$ and magnetic field $h$ is taking in Boltzman''s constant scaling i.e. Boltzmann's constant is
set to be $k_B = 1$.

The results of the previous section can be used for investigation
of the magnetic properties of our model.  The
magnetization of  arbitrary site is defined as
\begin{equation}\label{magnetization} m_\upsilon=\frac{Tr( S_\upsilon e^{ - H/T})}{Z}
\end{equation}
$S_\upsilon$   corresponding  spin operator
on sublattice $(\upsilon)$,
$H$ is the Hamiltonian (\ref{Htrial}) with   constants (\ref{gamma2}) and $Z$ is partition function
of the system. But according to (\ref{gamma2})
the  Hamiltonian of sublattice $(a)$ depends on $m_b$ through $\gamma_a$
and vice versa.
For defined above magnetization
we obtain
the following expression:
\begin{equation} \label{magnet}
m_a=\frac{1}{6}\cdot\frac{3\,\text{sinh}\left(\frac{3\gamma_a}{T}\right)+\text{sinh}\left(\frac{\gamma_a}{T}\right)+2e^{\left(\frac{6\lambda}{T}\right)}\text{sinh}\left(\frac{\gamma_a}{T}\right)}{\
\  \text{cosh}\left(\frac{3\gamma_a}{T}\right)+\text{cosh}\left(\frac{\gamma_a}{T}\right)+2e^{\left(\frac{6\lambda}{T}\right)}\text{cosh}\left(\frac{\gamma_a}{T}\right)},
\text{ and }\gamma_a=\mathrm{h}-(\mathrm{J}_2-\mathrm{J}_3)
m_b.
\end{equation}
The dependance of
magnetization $m_a$ from external magnetic field $h$ can be found by solving
 the this recursive equation for each value of magnetic field $h$.

At relatively high temperatures the recursive equation has one stable solution
and therefore magnetization curves of  sublattices $(a)$ and
$(b)$ coincide (see Fig. 2(a)).  \begin{figure}[t]
$\begin{array}{cc}
 \includegraphics[width=200pt]{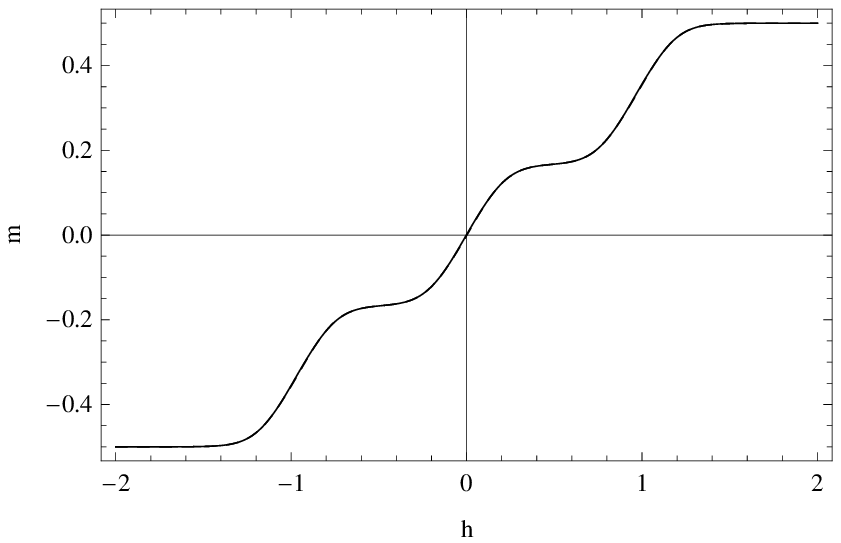}&  \includegraphics[width=200pt]{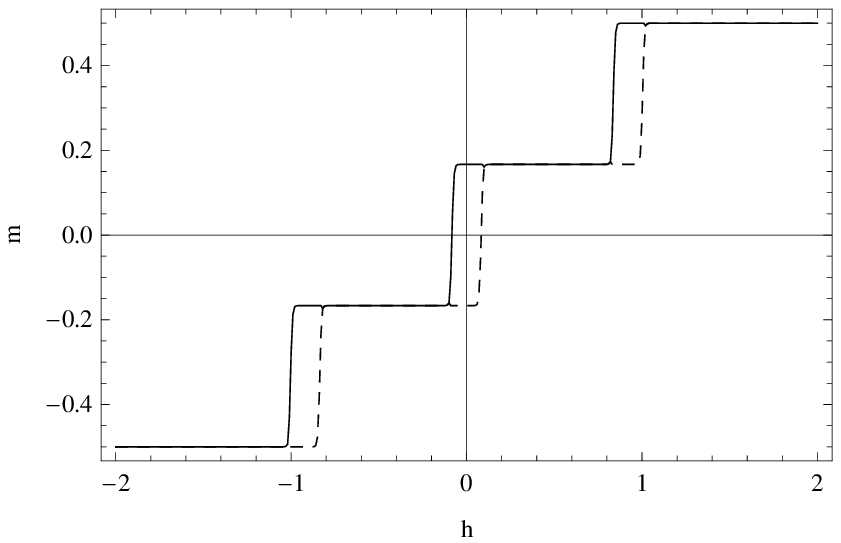}\\
 a)&b)
\end{array}$
\caption{  Magnetization $m_a$  versus external magnetic field $h$ for $J_2 = 3\ mK, J_{3} = 2.5\
mK.$ at a) T=0.15 mK b) T=0.01 mK.}
\end{figure}With decreasing temperature the solution of recursive equation ceases to be stable and, therefore, the magnetization of different sublattices are no longer equal. The partially
saturated phase emerges in form of the magnetization plateaus
(see Fig. 2(b) for  $T = 0.01\ mK$, $J_2 = 3\ mK, J_{3} = 2.5\
mK$), which can be associated with a staggered
magnetization or short range antiferromagnetism
(AF) in frustrated kagome geometry. Indeed, the appearance
of plateaus in magnetization curve at $m = \pm1/6$
can be explained as stability of trimeric  states in available ($\uparrow\uparrow\downarrow,\uparrow\downarrow\uparrow,
 \downarrow\uparrow\uparrow$) and ($\uparrow\downarrow\downarrow,\downarrow\uparrow\downarrow,\uparrow\downarrow\downarrow$) configurations. Moreover, zero field magnetisation of one sublattice becomes nonzero.\ In figure 3 the solid line shows the temperature dependence of the
magnetization in the absence of external magnetic field. The magnetization tents gradually to zero near the second-order transition temperature  $T_C$
between ordered $m_a\ne 0$ and disordered $m_a=0$ phases.
\begin{figure}[t]
\center \includegraphics[width=300pt]{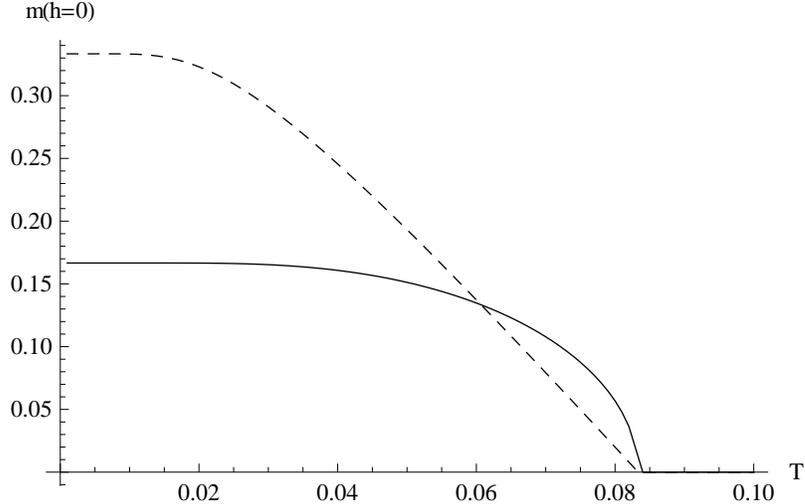}

\caption{  Dependence of the magnetization $ m_a $ (solid line) and
concurrence (dashed line)  at zero external field from temperature $ T $ at $ J_2 = 3 \;
\hbox {mK} $, $ J_3 = 2.5 \; \hbox {mK} $. }
\end{figure}
 \section{Concurrence and thermal entanglement}
The mean-field like treatment  transforms kagome
lattice to the set of noninteracting triangles
in effective field, therefore quantum correlations can be exactly
accounted. This allows in
terms of three-qubit XXX Heisenberg model in effective
magnetic field $\gamma$ to study  the thermal entanglement properties.   We will study
 the concurrence as a measure of pairwise entanglement \cite{wooters}. The concurrence
$C(\rho)$ corresponding to the density matrix $\rho$ is defined
as
\begin{equation}\label{18}\nonumber
C(\rho)=max\{\lambda_1-\lambda_2-\lambda_3-\lambda_4,0\},
\end{equation}
where $\lambda_i$ are the square roots of the eigenvalues of the
operator
\begin{equation}\label{19}\nonumber
\tilde\rho=\rho_{12}(\sigma_1^y\otimes\sigma_2^y)\rho^*_{12}(\sigma_1^y\otimes\sigma_2^y),
\end{equation}
where $\rho_{12}=Tr_3\rho$ is the
reduced density matrix of the pair and $\rho$ is defined in the following way
\begin{equation}
\rho=\frac{1}{Z}\sum\limits_{i=1}^8e^{-\frac{E_i}{T}}|\psi_{i}\rangle\langle\psi_{i}|,
\end{equation}
where $Z$ is the partition function of the system and $\psi_i$ and $E_i$
are eigenvectors and eigenvalues of the Hamiltonian $H_0^{(i)}$
respectively (see Eq. \ref{Htrial}).     $\rho_{12}$ has the following form
\begin{equation}\label{21}\nonumber
\rho_{12}=
\left(\begin{array}{cccc}
u&0&0&0\\
0&w&y&0\\
0&y&w&0\\
0&0&0&v\\
\end{array}\right),
\end{equation}
where $u,w,y$ and $v$ are some functions of variables $\gamma,\lambda$ and $T$. Using (\ref{18}),(\ref{19}) and (\ref{21})  one can find the following expression for  the concurrence $C(\rho)$
\begin{equation}\label{concur}
C(\rho)=max\{|y|-\sqrt{uv},0\}.\nonumber
\end{equation}
In this equation one must replace $\gamma$ with
$\mathrm{h}-2(\mathrm{J}_2-\mathrm{J}_3) m$ (see Eq.(\ref{gamma2})),
therefore the  concurrence $C(\rho)$ is the function of
magnetisation $m$. To calculate concurrence one must solve
transcendental equation (\ref{magnet}) for each set of parameter
values ($J_2,J_3,h,T$) and insert corresponding solution to the
equation (\ref{concur}).

\begin{figure}[t]$
\begin{array}{cc}
\includegraphics[width=210pt]{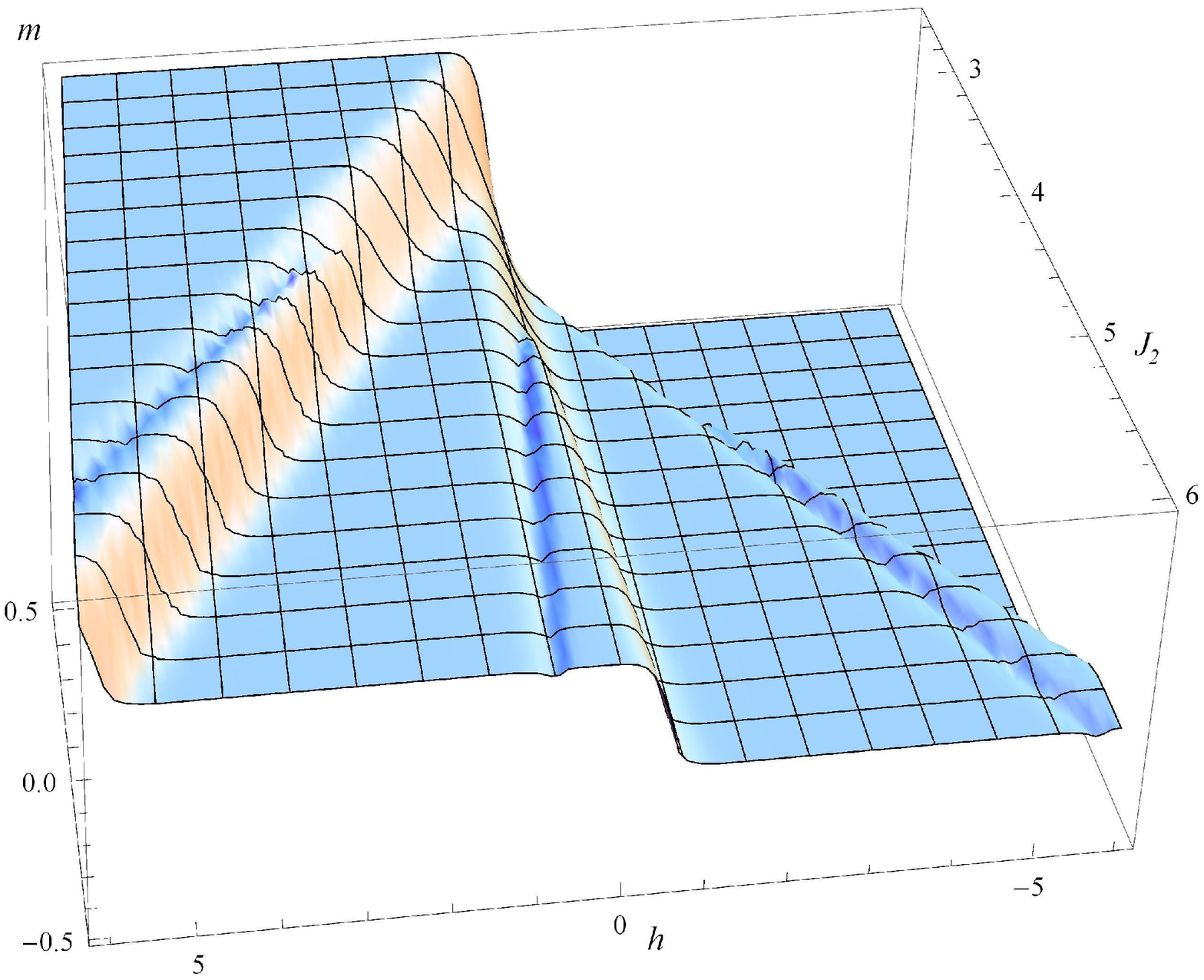}&\includegraphics[width=210pt]{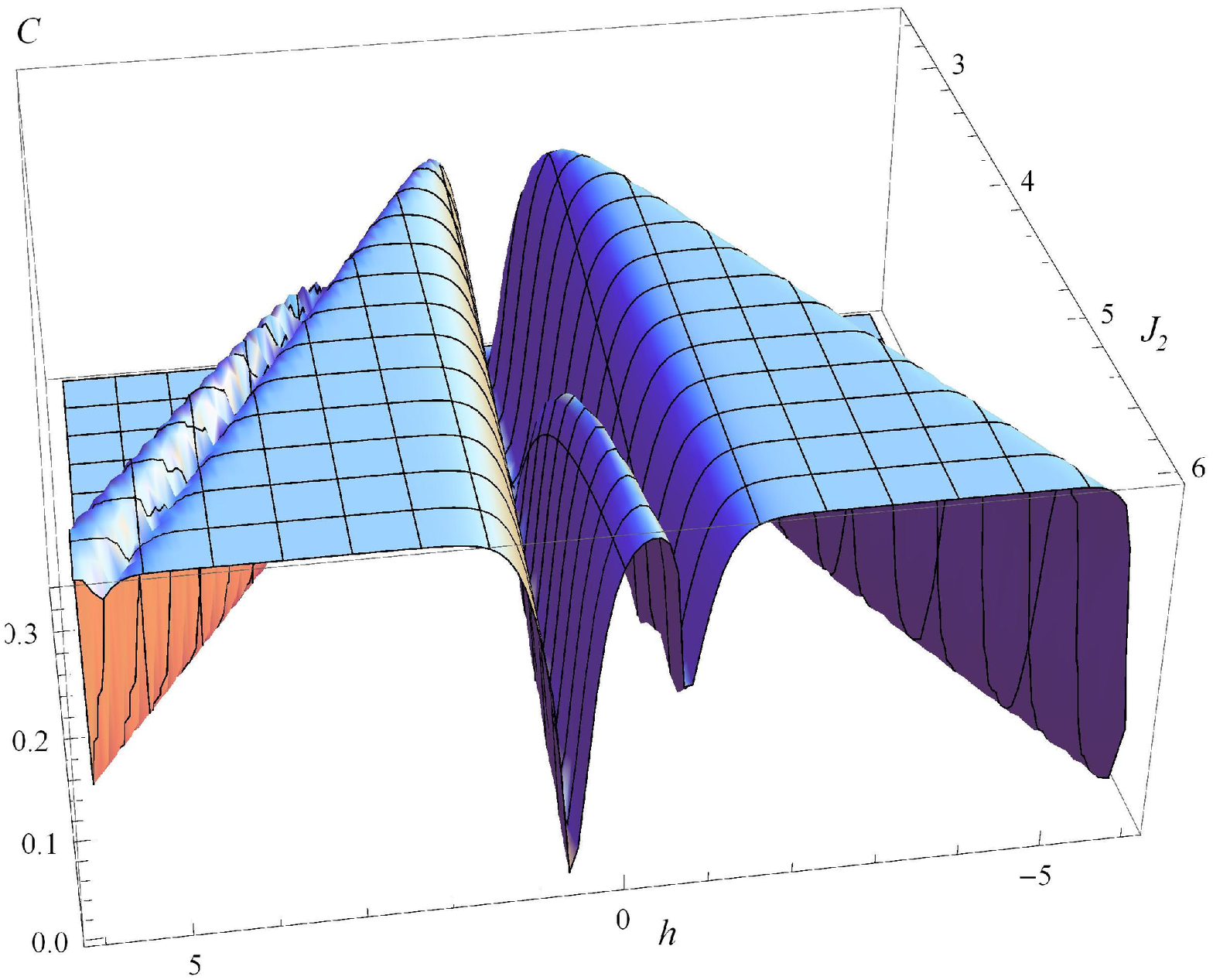}\\
a)&b)
\end{array}$
\caption{ Dependence for (a) magnetization $m$ and (b)
concurrence $C(\rho)$ versus the magnetic field $h$ and the coupling
constant $J_2$ at $J_3 = 2.5\ mK$ and $T =0.2\ mK.$ }
\end{figure}

It is curious to  discuss some similarities of statistical and
quantum characteristics of our system. We consider magnetization
 as a statistical characteristic. In figure
3(a) plotted the magnetization as a function of the coupling
constant $J_{2}$ (for fixed value of $J_3=2.5\ mK$) and the external
field $h$, at a relatively high temperature $T = 0.2\ mK$. As a
quantum characteristic we consider entanglement (concurrence
$C(\rho)$).   In figure 3(b) the  concurrence  as a function of the
$J_{2}$ ($J_3=2.5\ mK$) is shown for the same value of temperature.
Our calculations show that the magnetic characteristics is similar
to that of bipartite entanglement. Indeed, comparison of figures
3(a) and 3(b) shows that regions corresponding to the
 magnetization   plateaus, coincide with the plateaus on concurrence plot.
 \section{Conclusion}
 In this paper we find strong correlations between magnetic properties and quantum
entanglement in the Heisenberg model with two-, and three-site  exchange interactions in strong magnetic field on the  kagome lattice, which
correspond to the third layer of
fluid  $^3He$ absorbed on the surface of graphite.  We adopted variational mean-field-like
treatment (based on the Gibbs-Bogoliubov inequality) of separate clusters in effective
magnetic fields and studied  magnetic properties and concurrence as a measure of pairwise thermal entanglement. The system exhibits different magnetic behaviors,
    depending on the values of the exchange parameters ($J_2,  J_3$). We have obtained  the magnetization plateaus
    at low temperatures.
We have found, that in the  antiferromagnetic region behavior of the
concurrence  coincides with the magnetization one. The comparison of
magnetization and concurrence shows that regions corresponding to
the
 magnetization   plateaus, coincide with the plateaus on concurrence plot.
\section*{Acknowledgements} This work was supported by ECSP-09-08-SAS NFSAT  and  1981-PS, 2497-PS
ANSEF research grants.
–  


\end{document}